\documentclass[12pt,twocolumn]{article}

\usepackage{graphicx}
\usepackage{amsmath}
\usepackage{amssymb}
\usepackage{hyperref}
\usepackage{cite}
\usepackage{algorithm}
\usepackage{algpseudocode}
\usepackage{booktabs}
\usepackage{caption}
\usepackage{subcaption}
\usepackage{fancyhdr}
\usepackage{geometry}
\usepackage{booktabs}  
\usepackage{diagbox}   

\title{Machine Learning Approaches to Vocal Register Classification in Contemporary Male Pop Music}
\author{Alexander Kim, Prof. Charlotte Botha  \\
        Hamilton College}
\date{8/16/2024}

\begin{document}
\maketitle

\begin{abstract}
     For singers of all experience levels, one of the most fun and daunting challenges in learning, technical repertoire is navigating placement and vocal register in and around the passagio (passage between chest voice and head voice registers). Contemporary Pop and Musical Theater solos increasingly demand strong command through and above the first passagio, and the use of various timbre and textures to achieve a desired quality. Thus, it can be difficult to identify what vocal register within the vocal range a singer is using even for advanced vocalists. This paper presents two methods for classifying vocal registers in an audio signal of male pop music through the end-to-end analysis of textural features of mel-spectrogram images. Additionally, we will discuss the practical integration of these models for vocal analysis tools, and introduce a concurrently developed software called AVRA which stands for Automatic Vocal Register Analysis. Our proposed methods achieved consistent classification of vocal register through both Support Vector Machine (SVM) and Convolutional Neural Network (CNN) models, which shows promise for robust classification possibilities across a greater range of voice types and genre.   
\end{abstract}

\newpage 
\section{Introduction}

 In all voices, there are two muscles responsible for manipulating the vocal folds. The thyroarytenoid (TA) muscles shorten and thicken the folds while the cricothyroid (CT) muscles lighten and lengthen. TA dominant vocal functions result in 'chest voice' which the lower speech like vocal register in all genders. On the other hand, the CT muscles produce that pop falsetto head voice sound that has become prominent in the 21st century, and is often used to reach higher notes with a softer quality.

\subsection{Interpreting Mixed Voice in Pop Music}

 In vocal pedagogy, 'mixed voice' is used to denote the coordination of both TA and CT muscles, which imbues the sound with qualities from both chest voice and head voice. However, an experienced singer can change the balance of muscle function, so the term mixed voice covers the entire spectrum of chest-like to head-like timbres. For this reason, we divided the mixed voice classification into 'mix' and 'head mix' where mix denotes a more TA dominant, solid sound and head mix reflects more of the lighter head voice qualities. 
\\
\\
The challenge in identifying vocal register is the variety of factors involved from person to person. In male pop music, navigating the passagio demands control over mixed voice which raises the difficulty of vocal register identification. More formally, the passagio is sequence of pitches that bridge or transition between chest and head voice. Vocal registers don't have clear boundaries and can overlap in pitch particularly through the passagio. Besides pitch, other qualities like resonance, and placement play a role in determining the color and texture of the sound. This is how a singer can sing one note in two very different ways which is reflected in their choice of vocal register.

\subsection{Advantages of Data \\ Driven Vocal Analysis }

Expressing all these qualities of vocal audio with data was done by rendering samples of a single vocal register into mel-spectrograms. Thus, the proposed methods were achieved by analyzing the features and properties of mel-spectrograms. More specifically, the Fast-Fourier-Transform used to convert audio into spectrograms inherently increases the data from two to three dimensions. This enables for more robust automatic feature extraction via image classifiers, and improves data labeling through multi-modal identification of formants in audio clips and their corresponding images.
\\
\\
\begin{figure}[h]
\centering
\includegraphics[width=1 \columnwidth]{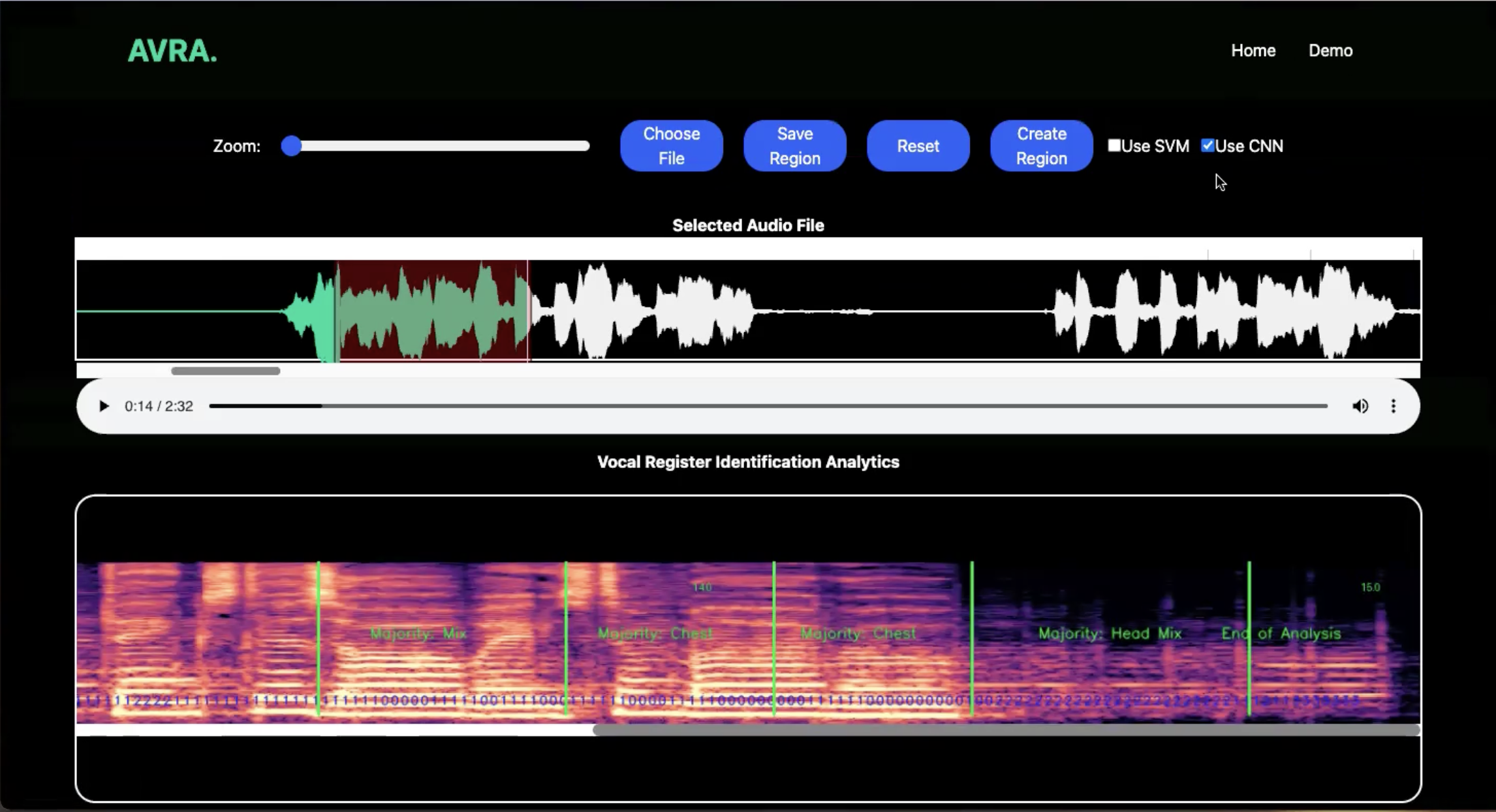}
\caption{AVRA analysis output example.}
\label{fig:AVAR_example}
\end{figure}
\\
\\
The Automatic Vocal Register Analysis (AVRA) software may help students and experienced singers further their understanding and practice of contemporary vocal technique. In AVRA, the user uploads an audio clips which is then processed into a sequence of mel-spectrograms. Then user then highlights the portion of audio they want analyzed and the software returns back the spectrogram with markers and labels for vocal register shifts. The temporal visual labeling of AVRA could help vocalists see exactly where they might be struggling, and what technique their target artist is using.

\section{Related Work}

\subsection{Audio Classification with Images}
Audio classification (particularly with deep learning) is accomplished by training models on spectrogram images. Identification of TA/CT muscle use has shown promising results with SVM models (Lacera, Mello, 2017). Since vocalizations are controlled by these muscles, we supposed that other qualities like vocal register could be identified with similar methods.

\section{Materials and \\ Methods}
This section describes the process and pipelines used for building the dataset, pre-processing images and extracting features, and model specifications and architectures. Finally, it will conclude by summarizing the training and evaluation process that determined which models were pushed to production. 
\subsection{Dataset}
Similarly to the paper referenced earlier, we used mel-spectrograms as the expression of the labeled audio clips. The lead vocals from WAV audio files of pop-songs were manually separated using the 'stem-splitter' tool in Logic Pro. Then, the raw audio was imported into Audacity and rendered into mel-spectrograms with a maximum frequency of 20000 Hz, gain of 20 dB, range of 80 dB, standard Roseus color scheme, and Hann window type as the default settings. These parameters were later replicated using the Librosa library in Python for model inference pre-processing. 
\\
\\
\begin{figure}[h]
\centering
\includegraphics[width=1 \columnwidth]{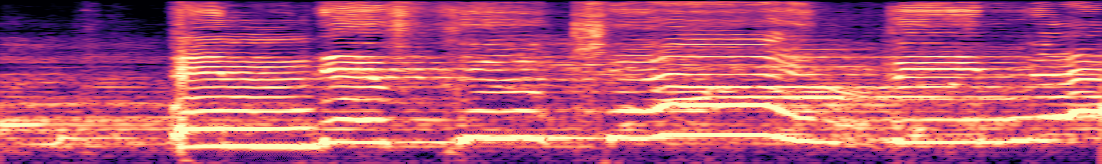}
\caption{mel-spectrogram of a 3-second vocal clip.}
\label{fig:example_spectrogram}
\end{figure}
\\
\\
Given that the dataset consists of 2010's pop music, pitches ranged from C3-C5, but were generally from A3-G4. Qualitative definitions for each vocal register were agreed upon at the beginning of the project with supplementation from various vocal technique texts (Peckham 2010). Then the raw vocals of each song were split into 3 second clips and rendered into spectrograms. An image was captured for any continuously used vocal register and then later processed into a standard size for the model inputs. These sub-images were given a one-hot encoded label for one of the 4 vocal registers. In these labels, 0 corresponds to Chest Voice, 1 corresponds to Mix, 2 corresponds to Head Mix, and 3 corresponds to Head Voice. A total of 1008 audio clips were rendered into mel-spectrograms which produced 7221 initial images before splitting and augmentation. 
\\

\subsection{SVM Classification}
Before experimenting with a CNN, we implemented a Support Vector Machine (SVM) for audio classification, utilizing spectrograms as input. This approach was motivated by the SVM's effectiveness in high-dimensional spaces by using the kernel trick to learn optimal separating hyperplanes between classes.
\\
\\
We began by pre-processing the spectrogram images. The input is standardized to 154x128 pixels, applying resizing and padding as necessary. To further generalize the dataset and address potential class imbalance, we used two data augmentation techniques: horizontal flipping and brightness adjustment. Each image was flipped across the Y-axis, and two additional versions were created by adjusting the brightness to 0.8 and 1.2 times the original. We limited our approach to these techniques to balance the benefits of a larger dataset against the constraints imposed by the super-linear scaling training time for SVM with a non-linear kernel.
\\
\\
The choice of these augmentation techniques was determined by the nature of spectrograms and the characteristics of vocal audio. In particular, horizontal flipping (instead of vertical flipping) preserves the frequency structure while providing reasonable variation in the time domain. This mimics potential variations in vocal phrasing without distorting the fundamental frequency relationships since time is represented on the x-axis and frequency on the y-axis. The brightness adjustments of factors 0.8 and 1.2 simulate variations in overall energy or volume of the vocal sample, which can occur due to factors like microphone distance, background noise, or singer intensity, without altering key spectral relationships. More extensive augmentation techniques, such as random rotations or vertical flips, were avoided as they would introduce unnatural spectral relationships that could confuse the model rather than enhance its learning.
\\
\\
After augmentation, the spectrograms were converted to grayscale, which matches the pre-processing technique Lacera and Mello used. The images were then flattened into one-dimensional vectors, resulting in a feature space of 19,712 dimensions (154x128). This transformation allowed us to represent each audio sample as a point in this high-dimensional space, where the SVM could effectively separate different vocal register classes. The combination of our targeted augmentation strategy and this high-dimensional representation provided the SVM with a rich, varied dataset that captured the characteristics of each vocal register while maintaining computational feasibility.
\\
\\
\begin{figure}[h]
\centering
\includegraphics[scale = 1]{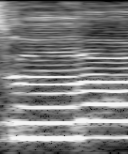}
\caption{pre-processed grey-scale spectrogram example of a balanced mix (label 1) spectrogram.}
\label{fig:grey_spectrogram}
\end{figure}
\\
\\
We chose a linear kernel for our SVM for computational efficiency, which was suitable given the high dimensionality of our feature space. The hyper-parameters were fine-tuned, with a notably small C regularization value of 0.0000025, promoting a larger margin and potentially better generalization. We also employed balanced class weights to mitigate any class imbalance issues in our dataset.
\\
\\
The SVM was trained to distinguish between four classes representing different vocal registers via a one-vs-rest strategy for this multi-class classification problem. This is the default approach in scikit-learn's SVM implementation.
\\
\\
To evaluate our model's performance, the dataset was split into training (80\%) and test (20\%) sets. We assessed the model using standard metrics including accuracy, precision, recall, and F1-score for each class. Additionally, we examined the confusion matrix to understand the model's performance across different vocal registers.
\\
\\
A key advantage of our SVM approach is its ability to provide probability estimates for each class, offering insights into the model's confidence in its predictions. This feature is particularly valuable in vocal register classification, where boundaries between classes can be subtle and gradual. By leveraging the SVM's strengths in handling high-dimensional data we trained a classifier capable of accurately distinguishing between different vocal registers based on spectrogram representations of audio samples.

\subsection{CNN Architecture}
The main goal of this project; however, was to employ a Convolutional Neural Network (CNN) for audio classification, leveraging its proven efficacy in image processing tasks. Like with the SVM, this approach was motivated by the transformation of audio signals into spectrograms, converting the audio classification problem into an image based task. The CNN's ability to capture hidden patterns in spectrograms, which often correspond to audio features such as harmonics, or formants in vowels, made it particularly suitable for this application.
\\
\\
Our CNN architecture consists of three convolutional layers followed by two fully connected layers. This structure was chosen to balance depth for feature abstraction with computational efficiency and generalization. The convolutional layers, provide translation invariance – a crucial property in audio classification where important features may occur at different time points or frequency bands within the spectrogram. Following each convolutional layer, we applied max pooling to reduce spatial dimensions, introducing a degree of position invariance and managing computational complexity.
\\
\\
The use of ReLU activation functions after each convolutional layer introduces non-linearity, allowing the model to learn, non-linear decision boundaries. This is essential for capturing the intricate relationships between different audio features. The final fully connected layers combined the high-level features extracted by the convolutional layers for classification, effectively learning the optimal weighting of various audio features to discriminate between classes.
\\
\\
This CNN architecture enables end-to-end learning, where both feature extraction and classification stages are learned simultaneously. This approach contrasts with traditional methods that often require separate feature engineering steps. Additionally, the hierarchical feature extraction in CNNs, combined with pooling operations, provides some robustness to noise and small variations in the input, which is a valuable characteristic in real-world audio classification tasks given imperfect sampling.
\\
\\
By employing this CNN architecture, we aimed to leverage these advantages for effective audio classification, aiming to outperform traditional ML methods that rely on laborious feature engineering or struggle with high dimensional data. The adaptability of this architecture to audio characteristics, its ability to handle variable-length inputs (with appropriate modifications), and its potential for capturing relevant features in many audio classification tasks motivated this choice.

\section{Results}
In this section, we will discuss the performance data from training the SVM and CNN models. We will discuss the effectiveness of both methods, with the CNN showing slightly superior performance. 
\subsection{Training Performance Tables}

\begin{table}[h]
\centering
\caption{SVM Performance Metrics - Training Set}
\label{tab:svm_performance_train}
\begin{tabular}{lccc}
\toprule
Class & Precision & Recall & F1-score \\
\midrule
Chest & 0.98 & 0.98 & 0.98 \\
Mix & 0.98 & 0.98 & 0.98 \\
HeadMix & 0.99 & 1.00 & 0.99 \\
Head & 1.00 & 1.00 & 1.00 \\
\midrule
Accuracy & \multicolumn{3}{c}{0.99} \\
\bottomrule
\end{tabular}
\end{table}

\begin{table}[h]
\centering
\caption{SVM Performance Metrics - Test Set}
\label{tab:svm_performance_test}
\begin{tabular}{lccc}
\toprule
Class & Precision & Recall & F1-score \\
\midrule
Chest & 0.93 & 0.94 & 0.93 \\
Mix & 0.93 & 0.91 & 0.92 \\
HeadMix & 0.92 & 0.97 & 0.94 \\
Head & 0.99 & 0.97 & 0.98 \\
\midrule
Accuracy & \multicolumn{3}{c}{0.94} \\
\bottomrule
\end{tabular}
\end{table}

\begin{table}[h]
\centering
\caption{Confusion Matrix for SVM Model (Test Set)}
\label{tab:confusion_matrix}
\resizebox{0.45\textwidth}{!}{%
\begin{tabular}{c|cccc}
\toprule
\diagbox{Actual}{Predicted} & Chest & Mix & HeadMix & Head \\
\midrule
Chest   & 1216 & 60   & 21   & 1    \\
Mix     & 82   & 1041 & 21   & 3    \\
HeadMix & 5    & 11   & 537  & 1    \\
Head    & 4    & 2    & 5    & 416  \\
\bottomrule
\end{tabular}
}
\end{table}

\subsection{SVM Results}
The SVM model demonstrated strong performance in classifying vocal registers, as shown by the metrics in Tables \ref{tab:svm_performance_train} and \ref{tab:svm_performance_test}. On the training set, the model achieved near-perfect classification across all registers, with training precision, recall, and F1-scores ranging from 0.98 to 1.00, and a training of 0.99.
\\
\\
On the test set, the model maintained robust performance, with a slight decrease compared to the training set. The overall accuracy on the test set was 0.94, with individual class metrics ranging from 0.91 to 0.99. The 'Head' register was particularly well-classified, with precision of 0.99 and recall of 0.97.
\\
\\
The confusion matrix (Table \ref{tab:confusion_matrix}) provides further insight into the model's performance. It reveals that misclassifications were most common between adjacent registers (e.g., Chest and Mix, or Mix and HeadMix). This suggests that the SVM successfully learned the features relevant to the labeled vocal registers and didn't just memorize its training data. More importantly, this reflects the continuous nature of vocal register transitions as a fundamental structure of it's understanding of the data. 

\subsection{CNN Results}

The CNN model demonstrated significant learning and generalization capabilities over the course of six training epochs. The training process showed a consistent decrease in loss and improvement in validation accuracy.
\\
\\
\begin{table}[h]
\centering
\caption{Training Performance Across Epochs}
\begin{tabular}{ccc}
\hline
\textbf{Epoch} & \textbf{Training Loss}  \\ \hline
1 & 1.394666  \\
2 & 0.764612    \\
3 & 0.548216  \\
4 & 0.302725  \\
5 & 0.116176  \\
6 & 0.064901  \\ \hline
\end{tabular}
\label{tab:training_performance}
\end{table}

\begin{table}[ht]
\centering
\caption{Validation Performance Across Epochs}
\begin{tabular}{ccc}
\hline
\textbf{Epoch} & \textbf{Val Loss} & \textbf{Val Accuracy} \\ \hline
1 & 0.781375 & 68.0\% \\
2 & 0.653277 & 72.3\% \\
3 & 0.445461 & 82.4\% \\
4 & 0.320593 & 88.5\% \\
5 & 0.185138 & 93.8\% \\
6 & 0.111533 & 96.2\% \\ \hline
\end{tabular}
\label{tab:test_performance}
\end{table}

In the first epoch, the model started with relatively high loss values, ranging from 1.394666 to 0.632968. The validation accuracy after the first epoch was 68.0\%, with a loss of 0.787375. As training progressed, we observed a steady improvement in both training loss and validation accuracy as shown in Tables \ref{tab:test_performance} and \ref{tab:training_performance}.
\\
\\
The training loss in the final epoch ranged from 0.064901 to 0.034674, showing a dramatic reduction from the initial epoch. This pattern of improvement in both training loss and validation accuracy suggests that the model effectively learned to extract relevant features from the mel-spectrograms while maintaining good generalization to unseen data. Furthermore, training was halted after the sixth epoch due to prevent overfitting following strong results on the previous two testing sets.

\section{Discussion}

So far, we demonstrated the efficacy of SVM and CNN machine learning, in classifying vocal registers in contemporary male pop music. Both models achieved high accuracy on their final validation sets, with the CNN slightly outperforming the SVM (96.2\% vs. 94\% test accuracy). However, the SVM has proven to be more practically useful in the AVRA (Automatic Vocal Register Analysis) software, potentially due to overfitting of the CNN. 
\\
\\
The SVM's performance, especially its ability to distinguish between adjacent registers, is noteworthy. The confusion matrix reveals that most misclassifications occur between adjacent registers (e.g., Chest and Mix, or Mix and HeadMix). This pattern aligns with the continuous nature of vocal register transitions and the inherent challenges in definitively categorizing these transitional sounds. Often times singers navigate difficult passages by smoothly transitioning between vocal registers, which shows the SVM's ability to distinguish between similar samples is commendable. The SVM's practical utility stems from its accuracy in deployment scenarios; however, the high dimensional feature space and size of the training set hinder its efficiency. Identifying a more efficient yet equally effective dataset could benefit user experiences with the model, particularly in the analysis of longer audio samples.  
\\
\\
While the CNN's learning curve is promising, showing rapid improvement across epochs, the SVM's superior generalization have made it more suitable for  adoption in various vocal analysis tools for male pop music specifically. The consistent decrease in both training and test loss for the CNN, coupled with increasing accuracy, suggests that the model successfully learned to extract relevant features from the mel-spectrograms but was prone to overfitting. However, the SVM's ability to generalize well to unseen data while maintaining a more straightforward implementation may be of benefit to real-world applications of vocal register classification with the current dataset. 
\\
\\
In both cases, our data augmentation techniques, particularly for the SVM, proved effective in enhancing model robustness and addressing potential class imbalance. The horizontal flipping and brightness adjustments simulated variations in vocal phrasing and overall energy without distorting fundamental frequency relationships, contributing to the models' ability to generalize.
\\
\\
Overall, both models' high performance validates our approach of using mel-spectrograms as input for vocal register classification. The success of this method suggests that mel-spectrograms effectively capture the timbrel and textural differences between vocal registers, allowing the models to learn discriminative features. However, with a much more training data, and classes to identify, the CNN may prove to be more effective because it has a shown propensity for relevant feature identification. Since the SVM does not scale as well to large datasets, this CNN model may adapt better to training data containing all voice types and a wider variety of contemporary music genres. With greater data training and quantity, a larger CNN has the potential to expand its domain into distinctly different types of music like classical singing, and across the full bass-soprano vocal fach system. 
\\
\\
Although the CNN slightly outperforms the SVM in overall accuracy, the SVM's practical advantages have made it the preferred choice in the concurrently developed AVRA software interface. In this application, musicians can upload their own vocal audio, render sections of the audio into spectrograms, and then apply either the SVM or CNN for automatic analysis. Output of labels in 10 pixel increments are marked with blue numbers (corresponding to the one-hot encoding scheme) at the bottom of the AVRA output as in figure \ref{fig:example_spectrogram}.
\\
\\
\begin{figure}[h]
    \centering
    \includegraphics[width=1\linewidth]{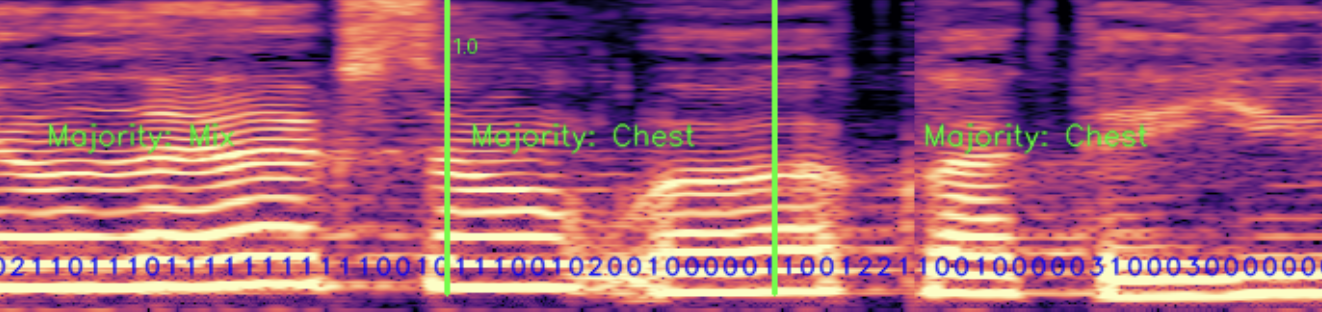}
\caption{example output of vocal audio analysis as shown on AVRA}
\label{fig:example_spectrogram}
\end{figure}
\\
\\
These results have significant implications for both music education and music production. In educational settings, the SVM-based system could be used to provide feedback to singers, helping them understand and refine their use of different vocal registers. In music production, this technology has been integrated into various tools for analyzing and categorizing vocal performances, aiding in tasks such as EQ tuning, or vocal effect application.
\\

\section{Conclusion}
In this paper, we have presented two machine learning approaches to vocal register classification that show great promise, achieving high accuracy while maintaining good generalization. The SVM has shown itself to be the preferred analysis option at the present moment, but the strengths of CNNs provide a solid foundation for this task, and show promise of a more robust tool that could be leveraged across a greater variety of vocal music. Additionally, we briefly discussed the potential for real-world applications and the development of AVRA for analysis of male pop music, and how this methodology could impact vocal pedagogy and music production, providing tools for both learning and creative applications in the realm of vocal performance. 
\subsection {Future Considerations}
Although the initial models are integrated into AVRA and performing well on this domain, it's important to note some limitations of this study. The dataset was limited to contemporary male pop music, and further research and data would be needed to generalize these findings to other genres or voice types. This would particularly benefited the CNN, and expand the target audience of the AVRA software. Additionally, while both models perform well on our dataset, real-world application would require testing on a wider range of singers and recording conditions to solidify performance. With this in mind, future work could explore several avenues:
\begin{enumerate}
    \item Further optimization of the pre-processing pipeline and SVM simplification for real-time audio processing to enhance immediate feedback capabilities.
    \item Expansion of the dataset to include more diverse vocal styles, genres, and voice types, particularly for transitioning or changing voices. 
    \item Investigation of the intractability of the model, particularly for the SVM, to understand which spectral features are most important for register classification.
    \item Comparison with other machine learning architectures, such as recurrent neural network or transformer variants, which might better capture temporal dependencies in the audio signal.
    \item Continued refinement of the AVRA application for singers and vocal coaches, leveraging its practical advantages.
\end{enumerate}
As we continue to refine these models and expand their applicability, we anticipate exciting developments at the intersection of machine learning and vocal music analysis. 

\section * {Acknowledgements}
We would like to thank the Emerson Foundation Grant Program for the support and funding of this project.


\begin{thebibliography}{99}

\bibitem{lacerda2017automatic}
Lacerda, E. B., \& Mello, C. A. B. (2017). \textit{Automatic classification of laryngeal mechanisms in signing based on the audio signal}. Procedia Computer Science.

\bibitem{peckham2010contemporary}
Peckham, A. (2010). \textit{The Contemporary Singer: Elements of Vocal Technique}. Hal Leonard Corporation.

\bibitem{shalev2014understanding}
Shalev-Shwartz, S., \& Ben-David, S. (2014). \textit{Understanding Machine Learning: From Theory to Algorithms}. Cambridge University Press.

\bibitem{peng2023voice}
Peng, Xiangyu., \&Xu, Huoyao., \& Liu, Jie., \& Wang, Junlang., \&He, Chaoming., (2023). \textit{Voice disorder classification using convolutional neural network based on deep transfer learning}. Scientific Reports

\bibitem{alzubaidi2021review}
Alzubaidi, L., Zhang, J., Humaidi, A. J., Al-Dujaili, A., Duan, Y., Al-Shamma, O., Santamaría, J., Fadhel, M. A., Al-Amidie, M., \& Farhan, L. (2021). \textit{Review of deep learning: concepts, CNN architectures, challenges, applications, future directions}. Journal of Big Data, 8(1), 53. https://doi.org/10.1186/s40537-021-00444-8

\end{thebibliography}
\end{document}